# GROUND-BASED MONITORING OF THE VARIABILITY OF VISIBLE SOLAR SPECTRAL LINES FOR IMPROVED UNDERSTANDING OF SOLAR AND STELLAR MAGNETISM AND DYNAMICS.

S. Criscuoli[1], L. Bertello[1], D. P. Choudhary[2], M. DeLand[3,4], G. Kopp[5], A. Kowalski[1,5], S. Marchenko[3,4], K. Reardon[1], A. A. Pevtsov[1], D. Tilipman[5]

[1]National Solar Observatory, 3665 Discovery Drive, Boulder, CO, USA - scriscuo@nso.edu;[2]San Fernando Observatory, California State University, Northridge, CA, USA; [3] Science Systems and Applications, Inc., Lanham, MD, USA; [4]Goddard Space Flight Center, Greenbelt, MD, USA; [5] Univ. of Colorado/LASP , 3665 Discovery Drive, Boulder, CO, USA.

## Synopsis

Long-term high-cadence measurements of stellar spectral variability are fundamental to better understand stellar atmospheric properties and dynamos, convective motions, rotational periods, eruptive phenomena, and stellar magnetospheres and winds. These, in turn, are fundamental for the detectability of exoplanets as well as the characterization of their atmospheres and habitability. The Sun, viewed as a star via disk-integrated observations, offers a means of exploring such measurements while also offering the spatially resolved observations that are necessary to discern the causes of observed spectral variations. In addition to solar variability studies, solar radiation also affects life on Earth and human activities in several ways. High-spectral resolution observations of the solar spectrum are fundamental for a variety of Earth-system studies, including climate influences, renewable energies, and biology.

The Integrated Sunlight Spectrometer (ISS) at SOLIS (Synoptic Optical Long-term Investigations of the Sun), has been acquiring daily high-spectral resolution (R~300 000) Sun-as-a-star measurements since 2006. It was designed to replace the decade-long National Solar Observatory programs at Kitt Peak and at Sacramento Peak. More recently, a few ground-based telescopes with the capability of monitoring the solar visible spectrum at high spectral resolution (R≳100 000) have been deployed (e.g. PEPSI, HARPS, NEID). However, the main scientific goal of these instruments is to detect exo-planets, and solar observations are acquired mainly as a reference. Consequently, their technical requirements are not ideal to monitor solar variations with high photometric stability, especially over solar-cycle temporal scales.

*The goal of this white paper is to emphasize the scientific return and explore the technical requirements of a network of ground-based spectrographs devoted to long-term monitoring of disk-integrated solar-spectral variability with* **high spectral resolution** *(R~100 000) and* **high photometric stability***, in conjunction with disk-resolved observations in selected spectral lines, to complement planet-hunters' measurements and stellar-variability studies.*

The proposed network of instruments offers the opportunity for a larger variety of *multidisciplinary* studies, which include improving our understanding of solar variability and its effects on the Earth's atmosphere, climate, and biological systems, which, in turn, offer a unique opportunity to understand how stellar magnetism affects stellar radiative emission and the properties of the planets that they host. All these topics are in scope of the Heliophysics 2024-2033 Decadal Survey statement of tasks.

# Introduction

Long-term high-cadence measurements of stellar spectral variability are fundamental to better understand stellar atmospheric properties and dynamos, evolution of the magnetic fields in stellar atmospheres, convective motions, rotational periods, eruptive phenomena, and stellar magnetospheres and winds. These, in turn, are fundamental for the detectability of exoplanets as well as the characterization of their atmospheres and habitability. The Sun, viewed as a star via disk-integrated observations, offers a means of exploring such measurements while also offering the spatially resolved observations that are necessary to discern the causes of observed spectral variations. Being the brightest and nearest star, the solar spectrum can be obtained with an extremely high signal-to-noise ratio and with very high spectral resolution. A few solar lines have been monitored for decades, using both space- (e.g. MgII, Ly$\alpha$) and ground-based (e.g. CaII H & K, HeI 10830) disk-integrated instrumentation at high spectral resolution. The Fourier Transform Spectrometer (FTS) at the McMath-Pierce Solar Telescope was capable of acquiring high-resolution (R~500 000) solar spectra across a wide spectral range. A few dozen visible lines were monitored with it from 1976 to 2010 (Livingston et al. 2007) on a monthly cadence. The program was replaced by the daily observations acquired with the Integrated Sunlight Spectrometer (ISS, Keller 2003) at SOLIS (R~300 000), which monitored nine spectral regions approximately 1 nm wide from 2006 until operations were halted in 2017 to relocate the instrument. Those observations are planned to resume at the end of 2022. Additionally, solar-spectral variations in broader spectral ranges have been observed (more or less) consistently over decades using space-based instrumentation, but these measurements typically encompass large spectral regions at coarse spectral resolution (less than ~0.1 nm, or R $\lesssim$ 10 000). A few ground-based telescopes with the capability of monitoring the solar visible spectrum at high spectral resolution (R$\gtrsim$100 000) have been deployed in the last few years (e.g., PEPSI, HARPS, NEID, EXPRES). However, the main scientific goals of most of these instruments are to study stars and exoplanets through high-precision radial-velocity measurements, and solar observations are acquired mainly as a reference. Consequently, their technical requirements (instrumental and operational) are not ideal to monitor solar variations with high photometric stability, especially over solar-cycle temporal scales. Moreover, not all the data produced by these instruments are publicly available.

*We propose long-term monitoring of disk-integrated solar-spectral variability with high spectral resolution and high photometric stability in addition to spatially resolved observations in select spectral lines obtained via a network of dedicated ground-based instruments.*

# Science Use Cases

There are several benefits of the proposed long-duration high-resolution solar spectral-line observations that will improve understanding of solar and stellar variability due to magnetic activity as well as the solar cycle and analogous stellar dynamos.

**Obtain Solar Reference Spectra for Various Activity Levels.** Several solar reference spectra exist but have discrepancies due to uncertainties in their absolute calibration, wavelength position, or contamination from telluric lines (Strassmeier et al. 2018). Typically, reference spectra with good absolute calibrations are composites of space-based, ground-based, and/or theoretical models (e.g. Meftah et al. 2018, Coddington et al. 2021), drawing on the high radiometric accuracies of space-based instruments and the superior spectral resolutions of ground-based measurements and models. These composites, especially when constructed to take into account different levels of magnetic activity, are employed in a large variety of fields, which include renewable energy (Myers 2017), calibration of remote-sensing instrumentation employed for Earth-atmosphere and -climate studies (e.g. Dobber et al. 2008, Pan and Flinn 2015), solar-system minor bodies (e.g. Morate et al. 2021), and marine biology (Frouin et al. 2018). Reference spectra also provide constraints for modeling the impact of solar radiation variability on the Earth's climate (Coddington et al. 2016). In the astrophysical context, solar reference spectra are essential for different science cases: development and validation of stellar atmospheric models, element-abundance estimates, modeling of stellar spectral irradiance variability, and studies of planetary atmospheres (Molaro and Monai 2012). It should be noted that a) differences between extant reference spectra in the visible and near-infrared are several percent on an absolute scale (see Fig. 1), whereas irradiance variability over the 11-year cycle is an order of magnitude smaller; b) several of these composite spectra make use of decades-old ground-based observations (usually data acquired with the McMath-Pierce FTS), which have high spectral resolution but include corrections for atmospheric absorption and sample very limited temporal periods; c) models of the solar spectrum are affected by uncertainties of the same order of magnitude (Criscuoli et al. 2020); and d) many of the observations contributing to these references are not temporally continuous, leading to gaps in solar-variability observations as well as offsets and trends between different measurements. Observed and modeled spectra of other stars are similarly affected by uncertainties and discrepancies (e.g. Cincunegui and Mauas 2004, Tilipman et al. 2021). Understanding and reducing such discrepancies for the solar-reference spectra will improve our knowledge of the spectral distribution of the Sun and other stars.

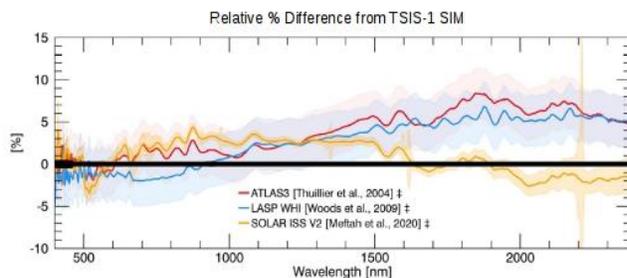

Fig. 1. Relative difference between different reference spectra and measurements acquired with TSIS-1 SIM radiometer. From Coddington et al. 2021.

*It is therefore of extreme importance that these Sun-as-a-star solar-spectral measurements be extended and the solar spectrum monitored continuously using modern, precise instrumentation.*
**Acquire continuous, low cadence (~daily) observations to understand variability on rotational and solar-cycle timescales.** This will address the following:
**Perform magnetometry via proxy line-measurements.** Almost exclusively, large-scale surveys of stellar magnetic activity rely on various proxies, such as broad-band photometry and line-activity indices (e.g., Ca II, Lyα, Mg II). In particular, variations of Ca II lines, especially H

& K, are highly correlated with variations of the stellar magnetic field (e.g., Pevtsov et al. 2016, Loukitcheva et al. 2009), and indices derived from Ca II K observations have served to monitor the long-term variability of the Sun (Petrie, Criscuoli, Bertello 2021) and other stars (Hall 2008). The advantage of using the Ca II line over other proxies is that it can be observed from the ground, making it relatively easy to construct long temporal records necessary to study stellar dynamos. *Accurate, continous Ca II measurements are our recommended highest-priority magnetic-proxy indicators*. The proposed instrument would allow us to continue monitoring of the fundamentally important Ca II lines, extending it to the transitions routinely used as the variability indicators in exo-planet searches (e.g., H$\alpha$, Na I, Gomes da Silva et al. 2011, Meunier et al. 2022), and explore new ones. Such homogeneous, long-term, high-cadence data should help in clarifying the yet elusive relationships between the surface magnetism and X-ray/EUV/UV emissions (Shoda and Takasao 2021).

**Correlate surface features and spectral lines**. Different lines are affected in different ways by the passage of magnetic structures. The relationships between the surface structures and line variability are complex and neither adequately documented nor well understood. Moreover, at times these relationships may not be reproduced in sufficient detail by the widely used solar models (Criscuoli et al. 2022). Comparison of disk-integrated solar spectra with spatially resolved images is crucial to validate this approach to estimate stellar magnetic-feature coverage. For instance, Thompson et al. 2020 recently compared disk-integrated HARPS-N measurements and HMI full-disk magnetograms to show that properties of a few Fe I, Ti II, and Mn I lines in the blue range (437-501 nm ) track the passage of faculae over the disk better than other indices, such as those derived from Balmer lines, which Marchenko et al. (2021) showed to be more sensitive to the passage of sunspots. This type of analysis serves two important science goals: a) *improve modeling of the contribution of magnetic features to stellar variability, which is essential for modeling the atmospheres of exoplanets through* **transmission spectroscopy** *(Rackham et al. 2022)*; b) *understand* **solar and stellar dynamos** *via continuous observations spanning several solar rotations and solar cycles.*

**Understand spectral-irradiance variability in the visible.** This portion of the spectrum largely reaches the Earth's surface, both playing a fundamental role for life (e.g. Egorova et al. 2018) and being relatively easy to monitor. As mentioned above, spectral-irradiance variations in this range are rather small, making them difficult to assess. Moreover, long-term observations are carried mostly from space with R~1000. This largely prevents adequate evaluation of the spectral-line contribution to the small – but crucially important for climate studies – irradiance variability in the visible. High spectral-resolution observations from the ground would integrate with low-resolution space-based measurements, allowing the construction of composites benefitting from both the high spectral-resolution ground-based measurements with good stability and absolute accuracies of the space-based measurements. This combination helps discern the contributions of high-resolution spectral lines to the solar variability observed at coarser spatial resolution, which, so far, has been estimated only through models, and the validation of spectral-irradiance models. It will also provide improved proxies, via correlations between spectral-line and spectral solar-irradiance variations in time.

**Better estimate UV and IR emissions.** UV radiation affects the Earth's atmosphere and climate via "top down" forcings (Haigh et al. 2010). Similarly, UV flux affects the atmospheres of exoplanets and their habitability. Due to atmospheric absorption, this short-wavelength radiation can only be observed and monitored from space-based platforms. These measurements cover only a few recent solar cycles and with somewhat discontinuous and inhomogeneous wavelength coverage and spectral resolution. For these reasons, substantial efforts have been dedicated to estimate solar and stellar UV emission and its variability through proxies and models (e.g., Youngblood et al. 2017). Ground-based observations are essential to ensure the construction of long-term, uniform estimates of UV radiation, as well as providing reliable UV proxy estimates that can be tied into the existing long-term proxy records.

Radiation at longer wavelengths also plays a fundamental role for climate and for life on Earth (e.g. Egorova et al. 2018). IR measurements are essential for the characterization of exoplanet atmospheres and for the identification of bio-markers. Like the UV, many IR spectral regions are largely absorbed by the Earth's atmosphere and thus require space-based observations. As in the UV, it would therefore be very useful to construct proxies of the long-wavelength IR flux that can be monitored with ground-based observations in order to provide solar- and stellar-variability estimates. The set of well-designed near-IR proxies may contribute to the rapidly expanding field of exoplanet studies that will assuredly receive a large boost from the recently launched JWST (an almost exclusively IR facility). *Our proposed effort will acquire the high spectral-resolution, long-duration measurements needed to provide such UV and IR proxies.*

**Acquire high-cadence observations of eruptive events and oscillations.**

**Flares.** There is increasing evidence that moderate- to high-intensity flares are associated with white light (WL) emissions, whose signatures are detectable in disk-integrated observations (e.g. Kretzschmar 2011, Woods et al. 2004). We thus expect both photospheric and chromospheric lines to react to intense flares. Observations of these variations, together with observations of the Balmer jump, would offer insight to understanding stellar flares, which can be both more frequent and energetic for young stars potentially hosting exoplanets (see the White Paper titled "Development of Integral Field Spectrographs to Revolutionize Spectroscopic Observations of Solar Flares and other Energetic Solar Eruptions"). Very high cadence (~ minutes) observations would be triggered by existing flare-watch programs and space-weather alerts.

**CMEs.** Stellar CMEs are detected by analyzing Doppler-shift and line-shape changes of spectroscopic lines, in particular Balmer lines (e.g., Vida et al. 2019). However, only a few events have been detected so far. Models suggest that CMEs could be detected in solar Balmer lines as well (Odert et al. 2020), although no observational detection has been reported yet. Like for flares, CME studies would be triggered by space-weather alerts.

**Oscillations.** Magnetic activity affects the amplitude and phase of solar oscillations. Several studies suggest this is the case for other stars as well (e.g., Santos et al. 2018), so that observations of oscillatory patterns can be employed to study the properties of the magnetic fields of other stars (e.g., Salabert et al. 2018, Bazot et al. 2018). Combined disk-integrated and spatially resolved observations of the Sun are essential to understand how disk-integrated observations of *p*-modes are affected by the presence of active regions on the disk and their properties (spatial distribution, size, magnetic field strength, etc.). Multiwavelength observations in particular are promising tools to study the effects of inclined magnetic fields on the phase of waves (see White Paper titled "Advancing the Understanding of Subsurface Structure and Dynamics of Solar Active Regions: An Opportunity with ngGONG"). Combined disk-integrated and full-disk, multi-wavelength observations would allow investigating how stellar inclination and magnetic-field distribution over the stellar surface affect stellar power spectra and hence inferences of their magnetic fields.

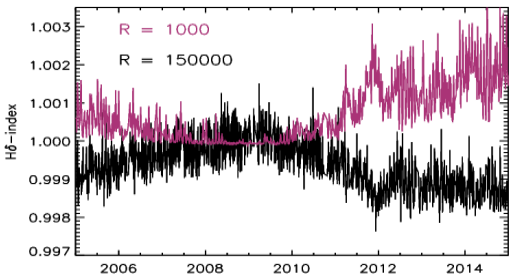

Fig.2 Modeled variability of H$\delta$ core-to-wing ratio index obtained degrading synthetic spectra to the resolution of spectrographs of different resolving power R. In both cases, the amplitude of the index variation over the decadal scale is small, at ~$10^{-3}$, but even the sign of the variation can change depending on the spectral resolution. This is likely due to the higher-resolution data correctly resolving line wings.

## Requirements

*High spectral resolution (R≳100 000) is required to investigate line-profile changes (Fig. 2).

*Photometric stability over decades of *at least* $10^{-3}$ to investigate variability over decadal temporal scales. See e.g. Fig. 2.

*Spectral range: contiguous 350 - 1100 nm (to include Ca II H&K and He I 10830 nm lines)

*Full-disk, spatially resolved images in select spectral lines (e.g., He I 10830, Ca II IR/UV, H$\alpha$)

*Cadence: ~ 6 hours to investigate variability over the solar-rotational timescale; ~10 seconds in flares/CMEs watch mode

*Duration: > 11 years to investigate variability at solar cycle (decadal) temporal scales

*Duty cycle: > 90 % for nearly continuous temporal coverage to study flares and CMEs; > 70 % for the other science cases. This will require a network of six ground-based observation sites for the former, high-cadence studies and at least three for the latter cases (Jain et al. 2021).

*Pristine sky. At least a single site, e.g., Mauna Loa, having occasional pristine sky is highly desirable for improved flux calibrations and to minimize the presence of telluric lines. This site can then provide cross-calibrations to other sites.

**Synergies with other current/future instruments**. In order to address the scientific goals listed above, it is fundamental that disk-integrated spectra are complemented with information provided by full-disk, spatially resolved observations in select spectral lines, as well as full-disk spectro-polarimetric observations in photospheric and chromospheric lines. This proposed combination of high spectral resolution disk-integrated measurements with spatially resolved ones is the synergy to enable physical understandings of underlying causal mechanisms that solar measurements provide that cannot be duplicated for (unresolvable) stellar observations. Such data are currently provided by facilities such as the ground-based SOLIS, GONG, PSPT, SFO, etc. and the space-based SDO, and in the future will be provided by instruments such as ngGONG and SPRING. Comparisons of the spectra acquired from the proposed network of spectrographs with observations acquired by planet-hunter instruments (e.g., HARPS-N, PEPSI, NEID, EXPRES) will be useful to address long-term instrumental effects and for the creation of composites, in a manner similar to what has been done so far with space-based solar irradiance measurements (e.g., Haberreiter et al. 2017). Comparison with radiometric spectral-solar irradiance (SSI) measurements obtained at much lower resolution will assess the contribution of spectral lines to SSI variability. Spectra acquired at different levels of solar activity will also be combined with radiometric SSI measurements to construct solar reference spectra (e.g., Coddington et al. 2021, Dobber et al. 2008) appropriate for different phases of the solar cycle.

## Proposed Instrumentation

A globally distributed network of at least *three* stations, each having an ***Echelle Spectrograph***, satisfies the requirements listed above. *Why a network?* An instrument network will increase the temporal coverage and longevity of the program (in case of failure of one or more instruments) and, via inter-calibration (see below), will ensure long-term measurement stability. We propose integrating the spectrographs with existing facilities to drastically reduce the costs of both the instrumentation and operations. We recommend Mauna Loa (for pristine sky), Cerro Tololo, and Udaipur among the (ng)GONG sites as excellent candidates for their geographical distribution. These three can provide > 70 % temporal coverage. To achieve the > 90 % duty cycle, at least three additional instruments located at other (ng)GONG sites or other observatories (e.g., San Fernando, Rome, IDA) may be added through collaborations with other foreign institutes.

**Measurement stability on decadal timescales is needed for discerning solar variability**. While absolute radiometry from the ground requires corrections for atmospheric transmission and is generally better achieved from space-borne platforms, relative stability measurements via our proposed approach of six ground-based stations has two advantages for maintaining long-term stability:

1. The instruments are accessible for regular calibration verification, component replacements and upgrades, and cleanings for contamination.
2. Simultaneous measurements from more than one station allow inter-calibrations between instruments for both signal levels and detections of drifts or other anomalous trends.

To achieve both **absolute accuracy and long-term stability**, we propose the following approach for the ground-based network:

1. Instruments are calibrated at both the component and system levels prior to deployment, such that each achieves accurate radiometric measurements.
2. *Atmospheric effects are corrected* via Langley extrapolation independently at each ground station.
3. *Periodic recalibrations* maintain and verify instrument accuracies. *Staggered recalibrations* allow the effects of any station's recalibrations to be discerned via inter-comparisons between sites that continued to operate during that site's downtime.

4. *Inter-comparisons between sites* transfer signal levels from the most accurate stations, providing relative cross-calibrations between all instruments. Expectations are that the Mauna Loa site will have the highest accuracies due to the pristine sky conditions, but at times, such as after an instrument upgrade or recalibration, another station may instead provide the highest levels of accuracy.
5. Stations having the lowest atmospheric-correction uncertainties will be *compared to space-based measurements to provide validations or corrections for radiometric accuracies*. These will be done in atmospherically clean windows after the higher-resolution ground-based spectral data are spectrally degraded to those of the space-based instruments. Such inter-comparisons can be timed around launches of pristine, new space-borne instruments and around periods of good atmospheric transmission for the ground-based stations.

This approach benefits from the advantages of both ground- and space-based measurements to achieve absolute accuracy and long-term stability.

## Costs

We estimate the cost for each instrument to be approximately US $300K (~120K for an off–the-shelf Echelle spectrograph, ~$130K for a 4Kx4K CCD, and ~$50K for integration and calibrations). The total cost of the network is therefore between $0.9M (for three instruments) and ~$1.8M (for six, with costs possibly shared with foreign institutes). This is in the project-cost range suitable for the Mid-Scale Innovations Program in Astronomical Sciences.

## Goals

During the period of the proposed effort, we plan to design, construct, and initiate operations. We also plan to: **a)** continue the NSO-ISS program, extend the available spectral ranges, and complement 'stellar' programs; **b)** use spectral-analysis tools to derive properties of lines; **c)** compare/integrate them with previous observations and existing ones to construct long-term time series of composites to employ for solar/stellar variability and solar-irradiance studies; **d)** construct reference spectra by combining observed high-resolution ground-based spectra with space-based spectral-irradiance observations; **e)** estimate the areal coverages of magnetic features extracted from full-disk observations; **f)** make the measurements and composites publicly available to the scientific community.

The proposed network of spectrographs offers the opportunity for a larger variety of **multidisciplinary studies which address the following Heliophysics 2024-2033 Decadal Survey statement of tasks**: 1) The structure of the Sun and the properties of its outer layers in their static and active states; 2) The consequences of solar variability on the atmospheres and surfaces of other bodies in the solar system and the physics associated with the magnetospheres, ionospheres, thermospheres, mesospheres, and upper atmospheres of the Earth and other solar system bodies; 3) Science related to the interstellar medium, astrospheres (including their stars), exoplanets, and planetary habitability.